\documentclass[fleqn,twoside]{article}
\usepackage{espcrc2}
\usepackage{epsfig}
\newcommand{\lsim}{\stackrel{<}{_\sim}}

\usepackage{graphicx}
\usepackage[figuresright]{rotating}

\newcommand{\AmS}{{\protect\the\textfont2
  A\kern-.1667em\lower.5ex\hbox{M}\kern-.125emS}}

\title{First hint of non-standard CP violation from $B \to \Phi K_S$ decay 
\\
\hspace{11.5cm}
{ \large{SLAC-PUB-9326\\
\hspace{13cm} July 2002}}
}

\author{Gudrun Hiller \address{
Stanford Linear Accelerator Center, Stanford University, Stanford, 
CA 94309, USA }
        \thanks{Work
supported by the Department of Energy, Contract
DE-AC03-76SF00515}}
       
\begin{document}

\begin{abstract}
We comment on the implications of the recently 
measured CP asymmetry in $B \to \Phi K_S$ decay. The data
disfavor the Standard Model at 2.7 $\sigma$ and -if the trend persists
in the future with higher statistics - require the existence of 
CP violation beyond that in the CKM matrix.
In particular, the $b \to s \bar s s$ decay amplitude would require new 
contributions of comparable size to the Standard Model ones with an order
one phase. 
While not every model can deliver
such a large amount of CP and flavor violation, 
those with substantial FCNC couplings to the $Z$ can 
reproduce the experimental findings.

\vspace{1pc}
\end{abstract}

\maketitle

\section{INTRODUCTION}

The breakdown of CP symmetry in the $b$-system has been established
from measurements of time-dependent asymmetries in
$B \to J/\Psi K$ \footnote{Throughout this paper 
$J/\Psi$ stands for all $c\bar c$ states included in the experimental
analyses for $\sin (2 \beta (J/\Psi K))$.} 
decays \cite{sin2betababar,sin2betabelle}.
In the  Standard Model (SM) the phenomenon of CP violation originates from
the CKM three generation quark mixing matrix \cite{ckm}. 
It is an impressive success of this CKM picture of CP and flavor violation
that the world average of the asymmetry in $B \to J/\Psi K$ 
decays
\cite{yossi}
\begin{equation}
\label{eq:avePsi}
\sin (2 \beta (J/\Psi K_{S,L}))_{world-ave}=+0.734 \pm 0.054
\end{equation}
agrees with the value extracted from experimental constraints 
from very different
processes such as those in the Kaon sector,
$\sin (2 \beta (J/\Psi K))_{fit}=+0.64 \ldots +0.84$ at
$95 \%$ C.L.~\cite{CKM-fit}.
However, 
this CKM paradigm is now challenged by the recently
reported measurements of CP asymmetries in $B \to \Phi K_S$ decays
by BaBar \cite{babarphi} 
\begin{eqnarray}
\label{eq:babar}
\sin (2 \beta (\Phi K_S))_{BaBar}=-0.19 ^{+0.52}_{-0.50} \pm 0.09 
\end{eqnarray}
and Belle \cite{bellephi}
\begin{eqnarray}
\label{eq:belle}
\sin (2 \beta (\Phi K_S))_{Belle}=-0.73 \pm 0.64 \pm 0.18
\end{eqnarray}
with resulting error weighted average
\begin{equation}
\label{eq:ave}
\sin (2 \beta (\Phi K_S))_{ave}=-0.39 \pm 0.41 \; .
\end{equation}
with errors added in quadrature.
The value in (\ref{eq:belle})
corresponds to the coefficient of the sine term in the
time dependent CP asymmetry, see e.g.~\cite{Anikeev:2001rk}.
Belle also quotes  a value for the direct CP asymmetry, i.e., ~the cosine term
$A_{\Phi K_S}=-0.56 \pm 0.41 \pm 0.12$ \cite{bellephi},
which is consistent with zero. In view of the current large experimental
uncertainties, we neglect direct CP violating effects on the decay
amplitudes in reporting the result of (\ref{eq:ave}). 
With increasing precision they
will become sensible and yield additional
information \cite{Fleischer:2001pc}.

In the SM the above decay modes are related such that 
the difference $D_{CP}$ of their asymmetries obeys 
\cite{Grossman:1996ke}-\cite{Grossman:1997gr}
\begin{equation}
\label{eq:diff}
D_{CP}\!=\! | \!\sin (2 \beta (\Phi K))- 
\sin (2 \beta (J/\Psi K)\!)| \!\lsim \! O(\lambda^2 \!)
\end{equation}
where $\lambda \simeq 0.2$ appears in Wolfenstein's parameterization of
the CKM matrix. 
Evaluation of (\ref{eq:avePsi}), (\ref{eq:ave}) yields
$D_{CP}=1.12 \pm 0.41$
hence violates the SM at 2.7 $\sigma$.
The impact of these experimental 
results on the validity of CKM and SM
is currently statistics limited.
Future prospects at the $B$-factories are
that the statistical error $\sigma_{\Phi K_S}(stat)$ 
can be expected to improve roughly by a factor
of three with an increase of integrated luminosity from
$0.1ab^{-1}$ to $1ab^{-1}$ \cite{Eigen:2001mk}
and it will take some time before we know $D_{CP}$ with sufficient 
significance to draw final conclusions.

In the following we entertain the possibility of 
a would-be measurement of $\sin (2 \beta(\Phi K_S))=-0.39$  or a similar
value which departs drastically from the SM expectation
of (\ref{eq:diff}).
We discuss the generic requirements to new physics (NP) models
to explain these values in Section \ref{sec:weak}.
In Section \ref{sec:new} we work out and discuss 
the reach of specific models in the observable $\sin (2 \beta(\Phi
K_{S,L}))$ and conclude in Section \ref{sec:end}.

\section{CONTRIBUTIONS TO $b \to s \bar s s$ FROM THE WEAK SCALE AND
BEYOND \label{sec:weak}}

\begin{figure}[htb]
\vskip 0.0truein
\centerline{\epsfysize=1.3in
{\epsffile{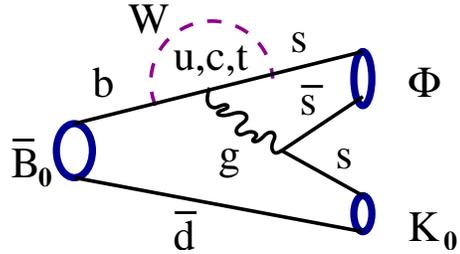}}}
\vskip -0.3truein
\caption{SM diagram contributing to $B \to \Phi K$ decay.}
\label{fig:SM}
\end{figure}

Time dependent measurements in $B_0, \bar B_0$ decays into a CP
eigenstate $f$ such as $J/\Psi K_S$, $\Phi K_S$ return 
the value of
$\sin (2 \beta(f))=\sin ( 2 \beta_{eff}+\triangle \Phi_f)$.
(As commented in the Introduction, we fix $|\bar A/A|=1$ 
to first approximation.)
Here, $\beta_{eff}$ is the phase from $B_0-\bar B_0$ mixing
and is common to all $B_0, \bar B_0 \to f$ decays
and $\triangle \Phi_f \equiv \arg(\frac{\bar A}{A})$ is the phase from
the decay amplitudes.
In the SM $\beta_{eff}= \beta$ and 
$\triangle \Phi_{J/\Psi K}$ and 
$\triangle \Phi \equiv \triangle \Phi_{\Phi K_S}\simeq
O(\lambda^2)$\cite{Grossman:1996ke}-\cite{Fleischer:2001cw}.
The ``golden-plated'' mode $B \to J/\Psi K$ is
mediated at the quark level by $b \to c \bar c s $ decay and 
receives a large contribution from tree level $W$ exchange. Hence, 
we expect
$\triangle \Phi_{J/\Psi K}$ to be subleading even in the presence of NP. 
On the other hand, the rare $B \to \Phi K$ decay appears in the SM
only at the loop level, see Fig.~\ref{fig:SM},
and therefore is generically more 
susceptible to (new) physics from the weak and higher scales.

Measurements of $\sin (2 \beta (\Phi K_S))$ and $\sin (2 \beta (J/\Psi
K_S))$ fix $\triangle \Phi$ up to a 4-fold ambiguity 
and in general have 8 pairs $(\beta_{eff}, \triangle \Phi)$ as solutions.
For example, let's take the good $O(10 \%)$ agreement between 
data on $\sin 2 \beta(J/\Psi K_{S,L}))$
and the SM fit for $\sin 2 \beta$  as an indication that
the value of $\beta_{eff}$ extracted is in the same branch as the
one from the SM fit, i.e.~we assume that $b \to c \bar c s $ decays
and $B_0-\bar B_0$ mixing
are dominated by the SM contribution.
(This concerns discrete
ambiguities and barring accidental cancellations 
does not affect our conclusions about large phases
in $b \to s \bar s s $ decays.)
Then, $\beta_{eff}=24^\circ$ and $\triangle \Phi=-70^\circ,-204^\circ$
using central values.
This requires a large source of CP violation in the
$b \to s \bar s s$ amplitude outside of the SM.
We recall that there is no conflict with a small direct CP asymmetry
as measured by BaBar
$A_{CP}(B^\pm \to \Phi K^\pm)=-0.05 \pm 0.20 \pm 0.03$
\cite{Aubert:2001rp}. While a large value for $A_{CP}$ would
inambiguously indicate the presence of NP, a small or vanishing one
could be caused by small or vanishing strong phases.

Lets illustrate what kind of scales could be invoked for an
interpretation of an $O(1)$ phase in the  $b \to s \bar s s$ decay amplitude.
The measured branching ratio  
${\cal{B}}(B_0 \to \Phi K_0)=(8.1 ^{+3.1}_{-2.5} \pm 0.8) \times 
10^{-6}$ \cite{PDG2002} is in 
agreement with the SM assuming factorization \cite{Ali:1998eb}, which
has however substantial errors from hadronic physics.
In the absence of a first principle precision calculation of hadronic 2-body
$B$ decays into light mesons we will not perform here a detailed study
of the $B \to \Phi K$  matrix element. 
Instead we assume that $b \to s \bar s s$ decays proceeds via a single 
flavor changing neutral current 
(FCNC) operator with appropriate Dirac structures $\Gamma_{i}$
\begin{equation}
\label{eq:O}
O=\xi_F g_X^2 \frac{\bar s \Gamma_1 b \bar s \Gamma_2 s}{M_X^2}
\end{equation}
generated from an interaction at scale $M_X$ with 
coupling $g_X$ and $\xi_F$ contains all flavor mixing information.
In the SM, $X$ is the weak scale, i.e.~$M_X=M_W$, $g_X=g_W$ and
$\xi_F =V_{tb} V_{ts}^*$ contains the CKM angles. The operator
contributes with Wilson coefficient $C_O$ renormalized at the $\sim  m_b$
scale of size of a few times $10^{-2}$ \cite{Ali:1998eb,BBL}. 
The NP contribution to $O$ has to be roughly 
of comparable size to the SM one to explain the
observed  $B_0 \to \Phi K_0$ branching ratio and has an order one CP phase in 
the overall mixing coefficient $\xi_F$ to explain a large CP asymmetry
induced by the  $b \to s \bar s s$ decay amplitude.

\begin{figure}[htb]
\vskip 0.0truein
\centerline{\epsfysize=1.3in
{\epsffile{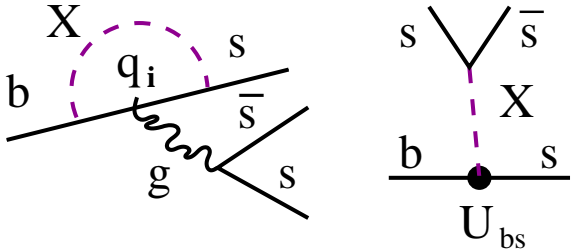}}}
\vskip -0.3truein
\caption{Examples of beyond the SM contributions to $b \to s \bar s s$ decays.}
\label{fig:BSM}
\end{figure}
Examples of contributions from physics beyond the SM to 
$b \to s \bar s s$ decays are shown in
Fig.~\ref{fig:BSM}.
The left diagram displays the effect of a new boson $X$ in the
FCNC loop with matter $q_i$ in close analogy to the SM mechanism.
If $g_X=g_W$, flavor angles $|\xi_F| =1$ and an 
$O(1)$ CP phase, and $C_{O}$ is SM-like, then this requires $M_X \simeq
400$ GeV to satisfy the conditions on size and CP breaking discussed above.
Assuming a larger 
Wilson coefficient of order 1 requires $M_X \simeq 2-3$ TeV.
Another possibility is tree level FCNC at the weak scale, 
where $M_X=m_Z$, $g_X=g_W$ and $\xi_F=U_{bs}$. 
This is illustrated in the right diagram of Fig.~\ref{fig:BSM}. 
The $sZb$ coupling has to be dominantly imaginary and
satisfy $|U_{bs}|\simeq 10^{-3}$ to be in the right ball park.

\section{WHICH NEW PHYSICS IN $B \to \Phi K$ ? \label{sec:new} }

In this section we examine the reach of different models
in the phase of the $b \to s \bar s s$ decay amplitude.
In particular we study the minimal supersymmetric Standard Model
(MSSM), a variant of the 2 Higgs doublet model (2HDM) III
which contains an extra source of
CP violation and a model with a vector-like down quark (VLdQ).
The CP reach in $b \to s \bar s s$ 
is estimated using the effective
Hamiltonian description and factorization \cite{Ali:1998eb,BBL}.
While this latter approach contains model dependence it gives the right 
pattern in which NP enters the rare decays.
Our findings are summarized in Table \ref{tab:pattern}.
Only those models with $\triangle \Phi \sim O(1)$ are able to
reproduce $\sin(2 \beta (\Phi K_S))=-0.39$ or a value
similarly different from $\sin(2 \beta (J/\Psi K_S))$.

What is the explanation in supersymmetry (SUSY) ?
To depart significantly from the SM with 
$\triangle \Phi \lsim O(\lambda^2)$
one has to go beyond the MSSM with minimal flavor violation (MFV), 
i.e.~allow for more CP and flavor violation than the one present in the SM
that is in the Yukawas.
Recall that gauge and anomaly mediation are MFV, whereas
in general SUSY GUTS \cite{Barbieri:1995rs} and effective SUSY models
\cite{Cohen:1996sq,Hisano:2000wy} are not.

Allowing for arbitrary mixing in the down squark sector,
the effect of gluino contributions in
$b \to s \bar s s$ decay has been analyzed 
in Refs.~\cite{Bertolini:1987pk,Ciuchini:1997zp,Lunghi:2001af}. As shown in 
\cite{Lunghi:2001af},
an order one NP contribution to the QCD penguins at the weak scale
can give at most a 10 $\%$ contribution at the scale $\mu \sim m_b$.
Imposing experimental 
constraints from $b \to s \gamma$  
a range $\triangle \Phi \lsim 0.7$ has been obtained \cite{Ciuchini:1997zp}.
A most important contribution in a generic MSSM without MFV
stems from up squark mixing between the second and third generation
which flips chirality, parametrized by $\delta^U_{23 LR}$.
This parameter is essentially unconstrained $|\delta^U_{23 LR}| \lsim
O(1)$, can be complex and induces an effective $sZb$ vertex
$|Z_{sb}| \lsim 0.1 |\delta^U_{23 LR}|$ 
defined as  \cite{GGG,Lunghi:1999uk}
\begin{eqnarray}
{\cal{L}}_{Z}=\frac{g^2}{4 \pi^2} \frac{g}{2 \cos \theta_W}
\bar b_L \gamma_\mu s_L Z_{sb}
\end{eqnarray}
These $Z$ penguins are constrained by $b \to s \ell^+ \ell^-$ decays
$|Z_{sb}| \leq 0.1$  \cite{GGG}-\cite{aghl}.
The contribution to $b \to \Phi s$  is then 
$\propto (-\frac{1}{2}+\frac{2}{3} \sin^2 \theta_W) 
\frac{g^2}{4 \pi^2} Z_{sb}$.
If the penguins are sizeable-
indicating the presence of large, complex up squark mixing in the
MSSM- they access values of $\triangle \Phi$ of  $O(1)$.
The effect of R-parity violating operators 
$\lambda^{\prime \prime}_{ijk} \bar u_i \bar d_j \bar d_k$ is small 
because there are no tree level contributions
to the $b \to s \bar s s$ operator (\ref{eq:O}) 
due to the symmetry properties of the super potential
\cite{Grossman:1996ke}.

We study the 2 HDM III as an example of a NP scenario with
an extended Higgs sector. 
The relevant model parameters are the
charged Higgs mass and the ``wrong'' Higgs couplings
of the third generation (we neglect all entries except the $(3,3)$
one) and their relative phase.
This new CP phase enters predominantly the
dipole operators such as the one 
with a gluon  $O_8 \simeq \bar s \sigma_{\mu \nu} b G^{\mu \nu}$,
see \cite{Bowser-Chao:1998yp} for details. 
This operator contributes to the $b \to s \bar s s$ amplitude,
though in the SM at subleading level compared to the QCD penguins, 
e.g.~\cite{Lunghi:2001af}. 
The 2HDM III model is constrained by non-observation of the charged
Higgs $m_{H^\pm}> 80$ GeV, the $b \to s \gamma$ branching ratio,
$B_0-\bar B_0$ mixing, the $\rho$ parameter and the neutron
electric dipole moment.
We scan over the allowed parameter space and obtain $\triangle \Phi \leq 0.2$.

A simple model beyond the SM with an enlarged matter sector is the one with
an additional vector-like down quark $D_4$.
The $(3 \times 4)$ dimensional extended CKM matrix $V$ includes 
mixing between  $D_4$ and the SM quark doublets 
and causes tree level FCNC couplings to the $Z$ \cite{Nir:1999mg}.
These are given as $U_{bs}=-V^d_{b4} V_{s4}^{d*}$ for $b \to s$
transitions, where $V^d$ diagonalizes the down sector.
This gives also the amount of CKM unitarity violation
$U_{bs}=\sum_{i=u,c,t} V_{ib}^* V_{is}$ which vanishes in the SM.
Following the discussion for the SUSY models with $Z$-penguins,
we relate $U_{bs} =-g^2/(4 \pi^2) Z_{sb}$
and  get $|U_{bs}| \lsim 10^{-3}$, 
slightly better than
the bound from \cite{Barenboim:2001fd}.
The reach of the VLdQ model in $\triangle \Phi$
is $O(1)$ in agreement with the estimates at the end of Section \ref{sec:weak}.
\begin{table*}[htb]
\caption{Reach of the SM and models beyond in $\triangle
\Phi=\arg(\bar A/A)$ in $b \to s \bar s s$ decays.}
\label{tab:pattern}
\renewcommand{\arraystretch}{1.3}
         \begin{tabular}{l|c|c|c|c}
\hline
\mbox{} & SM, MSSM with MFV
&generic MSSM & 2HDM III & VLdQ \\
\hline
$| \triangle \Phi | $
& ${\cal{O}}(\lambda^2)$ & ${\cal{O}}(1)$ &
         $\lsim 0.2$ & $ {\cal{O}}(1) $
         \\
\hline
         \end{tabular}
\end{table*}

\section{CONCLUDING REMARKS\label{sec:end}}

We have examined the
implications of the experimental results
\cite{babarphi,bellephi} on  
CP violation from interference between mixing and decay 
in $B \to \Phi K_s$ decays . These data are in conflict with 
the SM at $2.7 \sigma$ {\it and }
with many NP scenarios without $\triangle \Phi$ of $O(1)$, 
as compiled in Table \ref{tab:pattern}, such as the MSSM with MFV.
As we find, models with sizeable and complex 
$sZb$ couplings do have the required CP reach in
$b \to s \bar s s$ decays.
Note that anomalous couplings 
generically lead to large effects in the $sZb$ vertex
\cite{Burdman:1999fw}.
The $Z$-penguins contribute also to  
$b \to s \ell^+ \ell^-$ decays, $b \to s \nu \bar \nu$ decays
and $B_s-\bar B_s$ mixing \cite{GGG}.

A new CP violating NP contribution to the 
operator (\ref{eq:O}) will leak into other decays such as
$B\to K \eta, K \eta^\prime$ which do have a 
$s \bar s$ admixture.
Belle reported for the time-dependent asymmetry parameters
$\sin ( 2 \beta (\eta^\prime K_S))=0.76 \pm 0.36^{+0.05}_{-0.06}$ and
$A_{\eta^\prime K_S}=+0.26 \pm 0.22 \pm 0.03$
\cite{belleeta},\cite{bellephi}.
Due to the
anomalously large 
branching ratio of $B \to (K, X_s) \eta^\prime$  decays 
\cite{Ali:1998eb,etaprime}
the effect of NP in the $(\bar s b)(\bar s s)$ vertex
can be diluted in these channels by an enhanced SM contribution.
Hence, it is conceivable 
that $\sin ( 2 \beta (\eta^\prime K_S))$ is
closer to $\sin ( 2 \beta (J/\Psi K_S))$ than
$\sin ( 2 \beta (\Phi K_S))$
in agreement with the data and the
hypothesis of sizeable NP in $B \to \Phi K_S$ decays.
There might be as well already NP in the CP asymmetry in 
$B \to J/\Psi K_{S,L}$ decays (\ref{eq:avePsi}).
Excluding the possibility that NP in $b \to c \bar c s$ and/or
$B_0- \bar B_0$ mixing conspires such that the fit $\beta$ lives on a 
different branch than $\beta_{eff}$, this effect is at the 10 percent level.
Since ${\cal{B}}(B \to \Phi K)/{\cal{B}}(B \to J/\Psi K) 
\simeq 10^{-2}$ \cite{PDG2002} and assuming approximate flavor universality 
an order one NP contribution
to $B \to \Phi K_S$ is roughly a $10 \%$ correction to $B \to J/\Psi K$ 
which is within the errors.
Sensitivity to NP from measuring $\beta$ in different decays
is limited by the error on
$\sin 2 \beta_{fit}$, which can be improved if the error on 
$|V_{ub}|$ decreases and the SM background from $b \to u \bar u s$ 
contributions to $B \to \Phi K$ which has been suggested to bound 
by $SU(3)$ analysis \cite{Grossman:1997gr}.

\noindent
{\bf ACKNOWLEDGEMENTS}
It is a pleasure to thank David Atwood, Susan Gardner and Martin Schmaltz for 
stimulating discussions. I am grateful to Yossi Nir for communication.


\begin{thebibliography}{9}
\bibitem{sin2betababar} 
B.~Aubert {\it et al.}  [BABAR Collaboration],
Phys.\ Rev.\ Lett.\  {\bf 87}, 091801 (2001)
[arXiv:hep-ex/0107013];
%
B.~Aubert {\it et al.}  [BABAR Collaboration],
arXiv:hep-ex/0207042.

\bibitem{sin2betabelle} 
K.~Abe {\it et al.}  [Belle Collaboration],
Phys.\ Rev.\ Lett.\  {\bf 87}, 091802 (2001)
[arXiv:hep-ex/0107061];
T.~Higuchi  [Belle Collaboration],
arXiv:hep-ex/0205020.


\bibitem{ckm}
N.~Cabibbo,
Phys.\ Rev.\ Lett.\  {\bf 10}, 531 (1963).
%
M.~Kobayashi and T.~Maskawa,
Prog.\ Theor.\ Phys.\  {\bf 49}, 652 (1973).

\bibitem{yossi}
Y.~Nir, talk given at ICHEP 2002, Amsterdam, July 24-31, 2002.

\bibitem{CKM-fit} 
A.~H\"ocker, talk given at FBCP May 16-18, 2002, Philadelphia, USA.

\bibitem{babarphi}Aubert {\it et al.}(BABAR Collaboration), hep-ex/0207070.

\bibitem{bellephi} T.~Augshev, talk given at ICHEP 2002 (Belle
Collaboration), BELLE-CONF-0232; Abe {\it et al.}, 
BELLE-CONF-0201 hep-ex/0207098.

\bibitem{Anikeev:2001rk}
K.~Anikeev {\it et al.},
arXiv:hep-ph/0201071.

\bibitem{Fleischer:2001pc}
R.~Fleischer and T.~Mannel,
Phys.\ Lett.\ B {\bf 511}, 240 (2001)
[arXiv:hep-ph/0103121].



\bibitem{Grossman:1996ke}
Y.~Grossman and M.~P.~Worah,
Phys.\ Lett.\ B {\bf 395}, 241 (1997)
[arXiv:hep-ph/9612269].

\bibitem{Fleischer:1996bv}
R.~Fleischer,
Int.\ J.\ Mod.\ Phys.\ A {\bf 12}, 2459 (1997)
[arXiv:hep-ph/9612446].



\bibitem{London:1997zk}
D.~London and A.~Soni,
Phys.\ Lett.\ B {\bf 407}, 61 (1997)
[arXiv:hep-ph/9704277].

\bibitem{Grossman:1997gr}
Y.~Grossman, G.~Isidori and M.~P.~Worah,
Phys.\ Rev.\ D {\bf 58}, 057504 (1998)
[arXiv:hep-ph/9708305].

\bibitem{Fleischer:2001cw}
R.~Fleischer and T.~Mannel,
Phys.\ Lett.\ B {\bf 506}, 311 (2001)
[arXiv:hep-ph/0101276].



\bibitem{Eigen:2001mk}
G.Eigen {\it et al.}, 
in {\it Proc. of Snowmass 2001},
hep-ph/0112312.

\bibitem{Aubert:2001rp}
B.~Aubert {\it et al.}  [BABAR Collaboration],
Phys.\ Rev.\ D {\bf 65}, 051101 (2002)
[arXiv:hep-ex/0111087].


\bibitem{PDG2002} K.Hagiwara {\it et al.} (Particle Data Group), 
Phys.\ Rev.\ D {\bf 66}, 010001(2002).

\bibitem{Ali:1998eb}
A.~Ali, G.~Kramer and C.~D.~Lu,
Phys.\ Rev.\ D {\bf 58}, 094009 (1998)
[arXiv:hep-ph/9804363].

\bibitem{BBL} 
G.~Buchalla, A.~J.~Buras and M.~E.~Lautenbacher,
Rev.\ Mod.\ Phys.\  {\bf 68}, 1125 (1996)
[arXiv:hep-ph/9512380].

\bibitem{Barbieri:1995rs}
R.~Barbieri, L.~J.~Hall and A.~Strumia,
Nucl.\ Phys.\ B {\bf 449}, 437 (1995),
hep-ph/9504373;
K.~S.~Babu and J.~C.~Pati,
arXiv:hep-ph/0203029;
D.~Chang, A.~Masiero and H.~Murayama,
arXiv:hep-ph/0205111.

\bibitem{Cohen:1996sq}
A.~G.~Cohen {\it et al.},
Phys.\ Rev.\ Lett.\  {\bf 78}, 2300 (1997), 
hep-ph/9610252.

\bibitem{Hisano:2000wy}
J.~Hisano, K.~Kurosawa and Y.~Nomura,
Nucl.\ Phys.\ B {\bf 584}, 3 (2000),
hep-ph/0002286.

\bibitem{Bertolini:1987pk}
S.~Bertolini, F.~Borzumati and A.~Masiero,
Nucl.\ Phys.\ B {\bf 294}, 321 (1987).


\bibitem{Ciuchini:1997zp}
M.~Ciuchini {\it et al.},
Phys.\ Rev.\ Lett.\  {\bf 79}, 978 (1997)
[arXiv:hep-ph/9704274].


\bibitem{Lunghi:2001af}
E.~Lunghi and D.~Wyler,
Phys.\ Lett.\ B {\bf 521}, 320 (2001)
[arXiv:hep-ph/0109149].

\bibitem{GGG}
G.~Buchalla, G.~Hiller and G.~Isidori,
Phys.\ Rev.\ D {\bf 63}, 014015 (2001),
hep-ph/0006136.

\bibitem{Lunghi:1999uk}
E.~Lunghi, A.~Masiero, I.~Scimemi and L.~Silvestrini,
Nucl.\ Phys.\ B {\bf 568}, 120 (2000)
[arXiv:hep-ph/9906286].

\bibitem{ABHH}
A.~Ali {\it et al.}, 
Phys.\ Rev.\ D {\bf 61}, 074024 (2000),
hep-ph/9910221.

\bibitem{aghl} A.~Ali {\it et al.},
hep-ph/0112300, to appear in Phys.\ Rev.\ D.

\bibitem{Bowser-Chao:1998yp}
D.~Bowser-Chao, K.~m.~Cheung and W.~Y.~Keung,
Phys.\ Rev.\ D {\bf 59}, 115006 (1999),
hep-ph/9811235.

\bibitem{Nir:1999mg}
Y.~Nir,
arXiv:hep-ph/9911321.

\bibitem{Barenboim:2001fd}
G.~Barenboim, F.~J.~Botella and O.~Vives,
Nucl.\ Phys.\ B {\bf 613}, 285 (2001)
[arXiv:hep-ph/0105306].

\bibitem{Burdman:1999fw}
G.~Burdman, M.~C.~Gonzalez-Garcia and S.~F.~Novaes,
Phys.\ Rev.\ D {\bf 61}, 114016 (2000),
hep-ph/9906329.

\bibitem{belleeta}
K.~F.~Chen {\it et al.},
arXiv:hep-ex/0207033.

\bibitem{etaprime}
See, for example, 
D.~Atwood and A.~Soni,
Phys.\ Lett.\ B {\bf 405}, 150 (1997)
[arXiv:hep-ph/9704357];
A.~Ali, J.~Chay, C.~Greub and P.~Ko,
Phys.\ Lett.\ B {\bf 424}, 161 (1998)
[arXiv:hep-ph/9712372].
\end{thebibliography}
\end{document}